\documentclass[usenatbib]{mnras}
\usepackage{amssymb,amsmath}
\usepackage{graphicx}
\usepackage{xcolor,natbib}
\usepackage{multirow}
\usepackage[T1]{fontenc}
\usepackage{ae,aecompl}
\usepackage{newtxtext, newtxmath}
\usepackage{float}
\usepackage{subfigure}
\usepackage{longtable}
\usepackage{enumitem}
\usepackage{longtable,listings}
\usepackage[flushleft]{threeparttable}
\usepackage{parcolumns}
\def\oiii{[O~{\sc iii}]\ }
\def\nii{[N~{\sc ii}]\ }
\def\sii{[S~{\sc ii}]$\lambda6716$\AA\ }
\def\si{[S~{\sc ii}]$\lambda6731$\AA\ }

\def\oc{[O~{\sc iii}]$_c$}
\def\ob{[O~{\sc iii}]$_B$}
\def\oe{[O~{\sc iii}]$_E$}
\def\obj{SDSS J1547}
\def\o34{[O~{\sc iii}]$\lambda4364$\AA}

\title[SDSS J1547: different broad Balmer lines]
{SDSS J154751.94+025550 with double-peaked broad H$\beta$ but single-peaked broad H$\alpha$: 
a candidate for central binary black hole system?}

\author[Zhang X. G.]{XueGuang Zhang$^{1}$
\thanks{Contact e-mail: \href{mailto:xgzhang@njnu.edu.cn}{xgzhang@njnu.edu.cn}}\\
        $^{1}$School of Physics and Technology, Nanjing Normal University,
	        No. 1, Wenyuan Road, Nanjing, 210023, P. R. China}

\pubyear{2021}

\begin{document}

\label{firstpage}
\pagerange{\pageref{firstpage}--\pageref{lastpage}}

\maketitle

\begin{abstract} 
In this manuscript, an interesting blue Active Galactic Nuclei (AGN) SDSS J154751.94+025550 (=SDSS J1547) 
is reported with very different line profiles of broad Balmer emission lines: double-peaked broad H$\beta$ 
but single-peaked broad H$\alpha$. SDSS J1547 is the first AGN with detailed discussions on very different 
line profiles of the broad Balmer emission lines, besides the simply mentioned different broad lines in 
the candidate for a binary black hole (BBH) system in SDSS J0159+0105. The very different line profiles 
of the broad Balmer emission lines can be well explained by different physical conditions to two central 
BLRs in a central BBH system in SDSS J1547. Furthermore, the long-term light curve from CSS can be well 
described by a sinusoidal function with a periodicity about 2159days, providing further evidence to 
support the expected central BBH system in SDSS J1547. Therefore, it is interesting to treat different 
line profiles of broad Balmer emission lines as intrinsic indicators of central BBH systems in broad line 
AGN. Under assumptions of BBH systems, 0.125\% of broad line AGN can be expected to have very different 
line profiles of broad Balmer emission lines. Future study on more broad line AGN with very different 
line profiles of broad Balmer emission lines could provide further clues on the different line profiles 
of broad Balmer emission lines as indicator of BBH systems.
\end{abstract}


\section{Introduction}

    Broad emission lines from central broad line regions (BLRs) are fundamental characteristics of 
broad line Active Galactic Nuclei (AGN) \citep{sm00, gm09, kz13, vm20}. Due to limitations of modern 
observational techniques, central BLRs with distances of tens to hundreds of light-days \citep{kas00, 
ben13} to central black holes (BHs) cannot be directly spatial resolved. Detailed structures of central 
BLRs are determined through broad line emission features, especially long-term variabilities of broad 
emission lines. And then, different geometric structures of central BLRs have been reported, such as 
disk-like BLRs especially determined through double-peaked broad emission lines \citep{ch89, el05, 
st03, zh13, st17}, as structures determined through modeling the continuum variability and response in 
emission-line profile changes as a function of time \citep{gp13, gp17, df18, bd20}. 

   Among the broad emission lines in broad line AGN, broad H$\alpha$ and broad H$\beta$ are the two 
strongest optical recombination emission lines, such as the detailed emission line properties in composite 
spectrum of AGN and quasars as discussed in \citet{bt01, vr01}. Not similar as the commonly considered 
different emission regions between high-ionization broad lines and low-ionization broad lines \citep{bh16}, 
broad Balmer lines are well accepted to come from the totally same emission regions. Therefore, totally 
similar line profiles of broad Balmer emission lines are expected due to totally similar dynamical 
structures. And as the discussed results in large samples of broad line AGN and quasars in \citet{gh05}, 
similar line profiles of broad Balmer lines have been well reported, such as the reported strong linear 
line width correlation between broad Balmer lines in SDSS quasars. 

\begin{figure*}
\centering\includegraphics[width = 16cm,height=10cm]{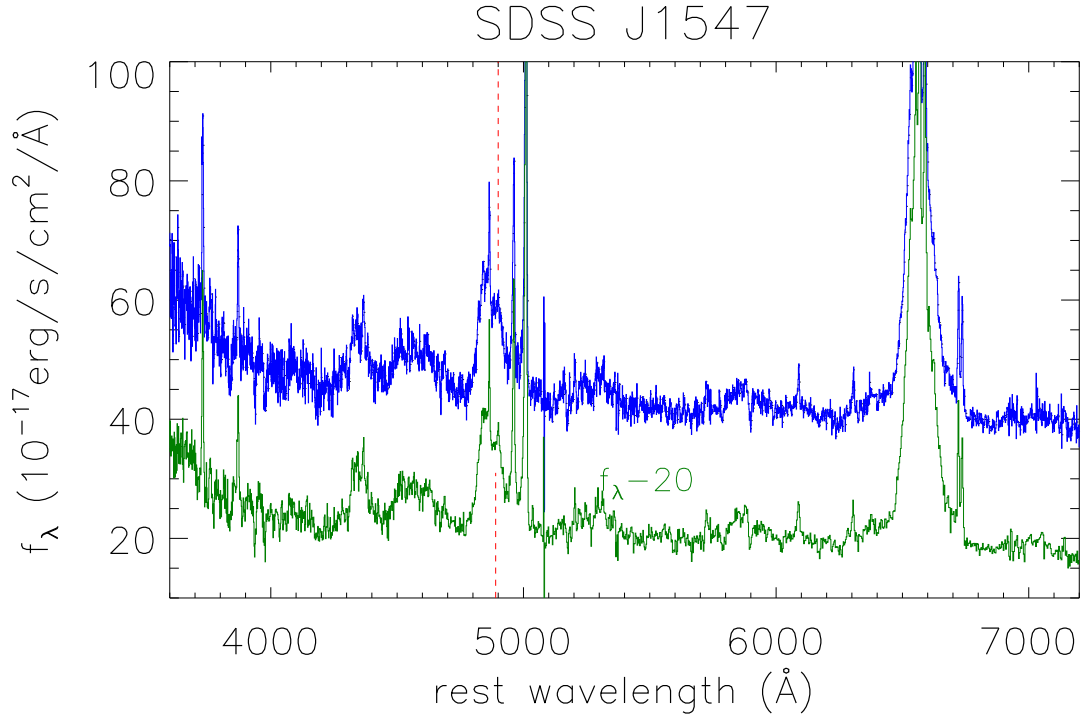}
\caption{SDSS spectra of SDSS J1547. Solid blue line shows the spectrum observed in MJD=52027, 
solid dark green line shows the spectrum observed in MJD=52045 with $f_\lambda$ minus 20. Vertical
dashed red line marks the second peak position in the broad H$\beta$.
}
\label{spec}
\end{figure*}

   However, if there were two central distinct BLRs with large distance enough for the broad Balmer 
emission lines, the observed broad Balmer emission lines should include two broad components from the 
two BLRs. Once there were different dust obscurations on the two BLRs or different physical local 
conditions (different ionization parameters, different electron densities, etc.) of the two BLRs, 
different line profiles of the observed broad Balmer emission lines could be expected due to the 
different circumstances around the two BLRs. The case of different obscurations is not easy to be 
imagined, unless the two BLRs have space distance very large enough. Whereas, the case of different 
physical local conditions can be well and commonly expected in AGN, especially in the well-known 
supermassive binary black hole (BBH) systems in AGN. BBH systems have been thought to be inevitable 
outcomes of the merger-driven galactic evolutions \citep{bb80, tw96, vh03, dm05, ca09, kp12, pl17, 
sb21}. And candidates for BBH systems have been reported in dozens of active galaxies through different 
techniques, especially based on properties of either spectroscopic properties or long-term variability 
properties combining with spatially resolved images, such as the reported candidates for BBH systems 
in \citet{zw04, rt09, sl10, gd15a, gm15, le16, dv19, kw20, lw21}. In a BBH system, each BH accreting 
system has a surrounding BLRs with dependent physical local parameters of ionization parameters, 
electron temperature, electron density, etc.. Therefore, different physical local conditions of the 
two BLRs could be common in BBH systems, and it is interesting to detect different line profiles of 
broad Balmer emission lines in broad line AGN harbouring BBH systems.

   In the manuscript, we report the robust and confirmed different line profiles of broad Balmer 
lines in the broad line AGN \obj: the double-peaked broad H$\beta$ and the broad H$\gamma$ but the 
single-peaked broad H$\alpha$, which could provide further clues on a central probable BBH system. 
In Section 2, the main spectroscopic results are shown. In Section 3, accretion disk origin 
of the broad Balmer emission lines are mainly discussed, providing evidence to rule out the accretion 
disk origin to explain the different line profiles of broad Balmer emission lines of \obj. In Section 4, 
BBH model is mainly discussed in \obj, confirmed by the detected optical quasi-periodic oscillations in 
the long-term variabilities of \obj. In Section 5, we simply discuss how many AGN similar as \obj~ 
can be found in SDSS (Sloan Digital Sky Survey). In Section 6, the main conclusions are given.
And in the manuscript, we have adopted the cosmological parameters of 
$H_{0}=70{\rm km\cdot s}^{-1}{\rm Mpc}^{-1}$, $\Omega_{\Lambda}=0.7$ and $\Omega_{\rm m}=0.3$.

\begin{figure*}
\centering\includegraphics[width = 16cm,height=20cm]{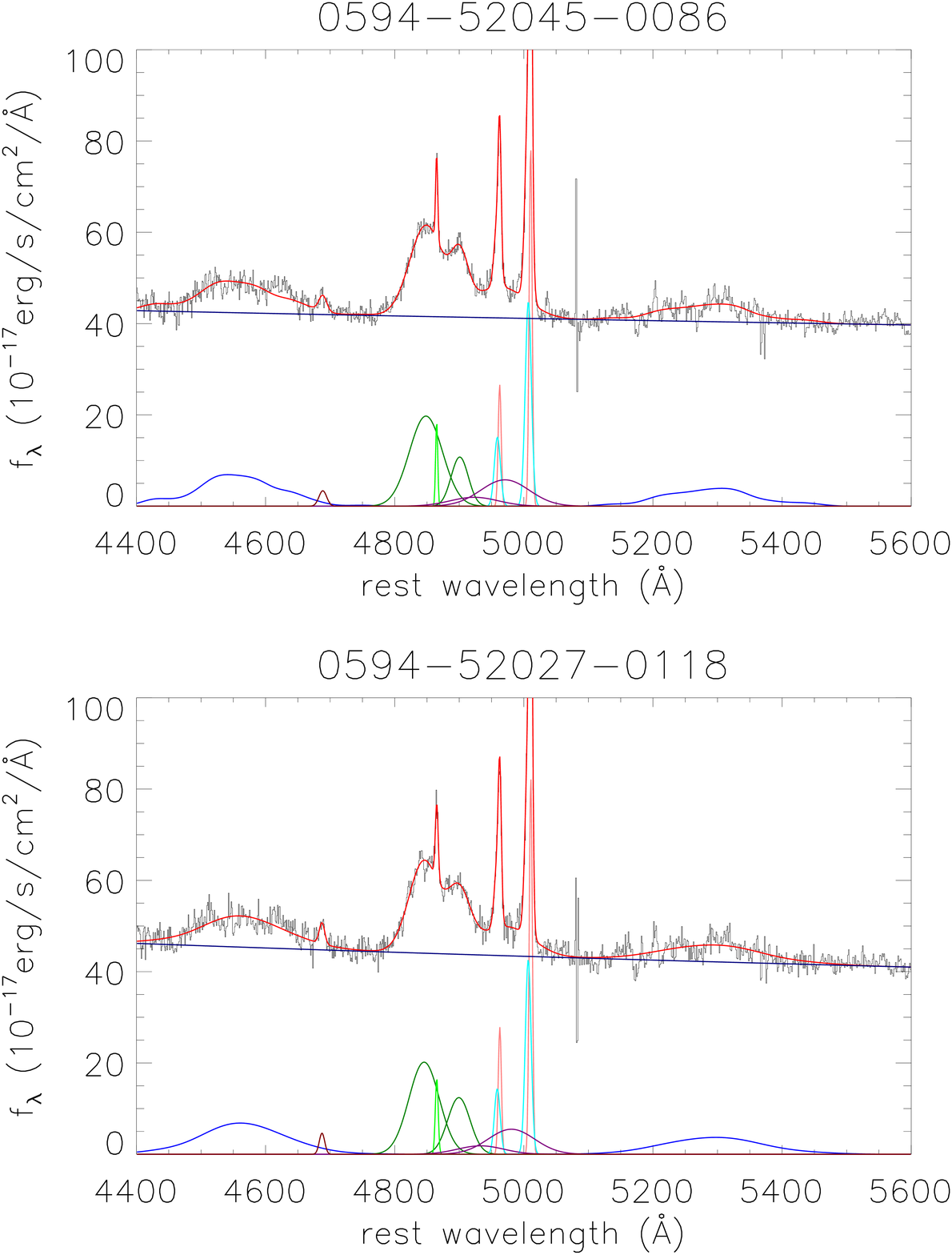}
\caption{Best descriptions to the emission lines around H$\beta$ in MJD=52045 (top panel) and in 
MJD=52027 (bottom panel). In each panel, solid black line and solid red line show the line spectrum 
and the determined best descriptions, respectively. Solid dark blue line shows the determined AGN 
continuum emissions, solid blue line shows the determined optical Fe~{\sc ii} emissions, solid dark 
green lines show the determined two broad Gaussian components in the broad H$\beta$, solid green 
line shows the determined narrow H$\beta$, solid pink and cyan lines show the determined core and 
broad \oiii components, solid purple lines show the determined extremely extended broad \oiii 
components, solid dark red line shows the determined broad He~{\sc ii} line. 
}
\label{hb}
\end{figure*}

\begin{figure}
\centering\includegraphics[width = 8cm,height=10cm]{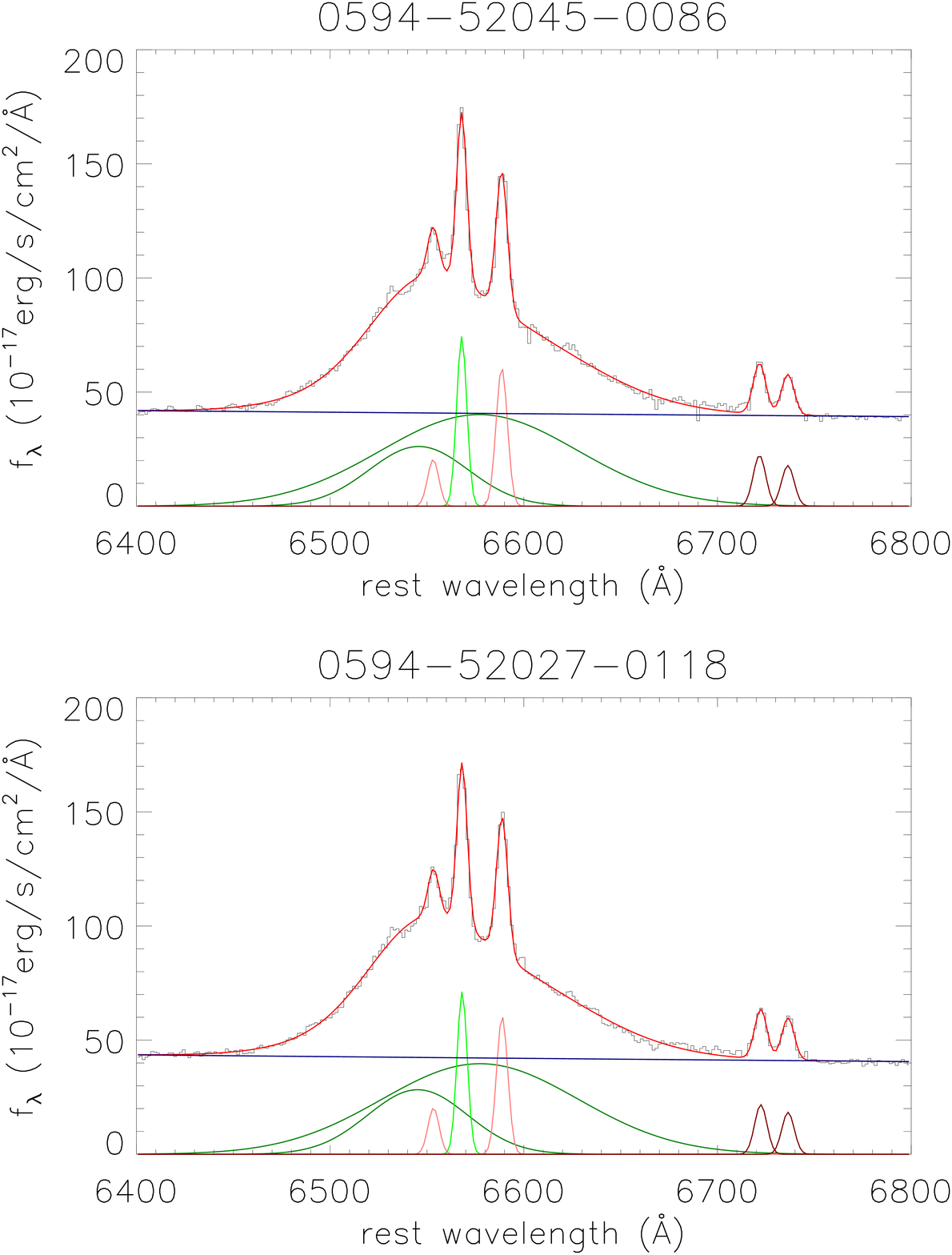}
\caption{Best descriptions to the emission lines around H$\alpha$ in MJD=52045 (top panel) and in 
MJD=52027 (bottom panel). In each panel, solid black line and solid red line show the line spectrum 
and the determined best descriptions, respectively. Solid dark blue line shows the determined AGN 
continuum emissions, solid dark green lines show the determined two broad Gaussian components in 
the broad H$\alpha$, solid green line shows the determined narrow H$\alpha$, solid pink lines show 
the determined [N~{\sc ii}] doublet, solid dark red lines shows the determined [S~{\sc ii}] doublet.
}
\label{ha}
\end{figure}

\begin{figure}
\centering\includegraphics[width = 8cm,height=10cm]{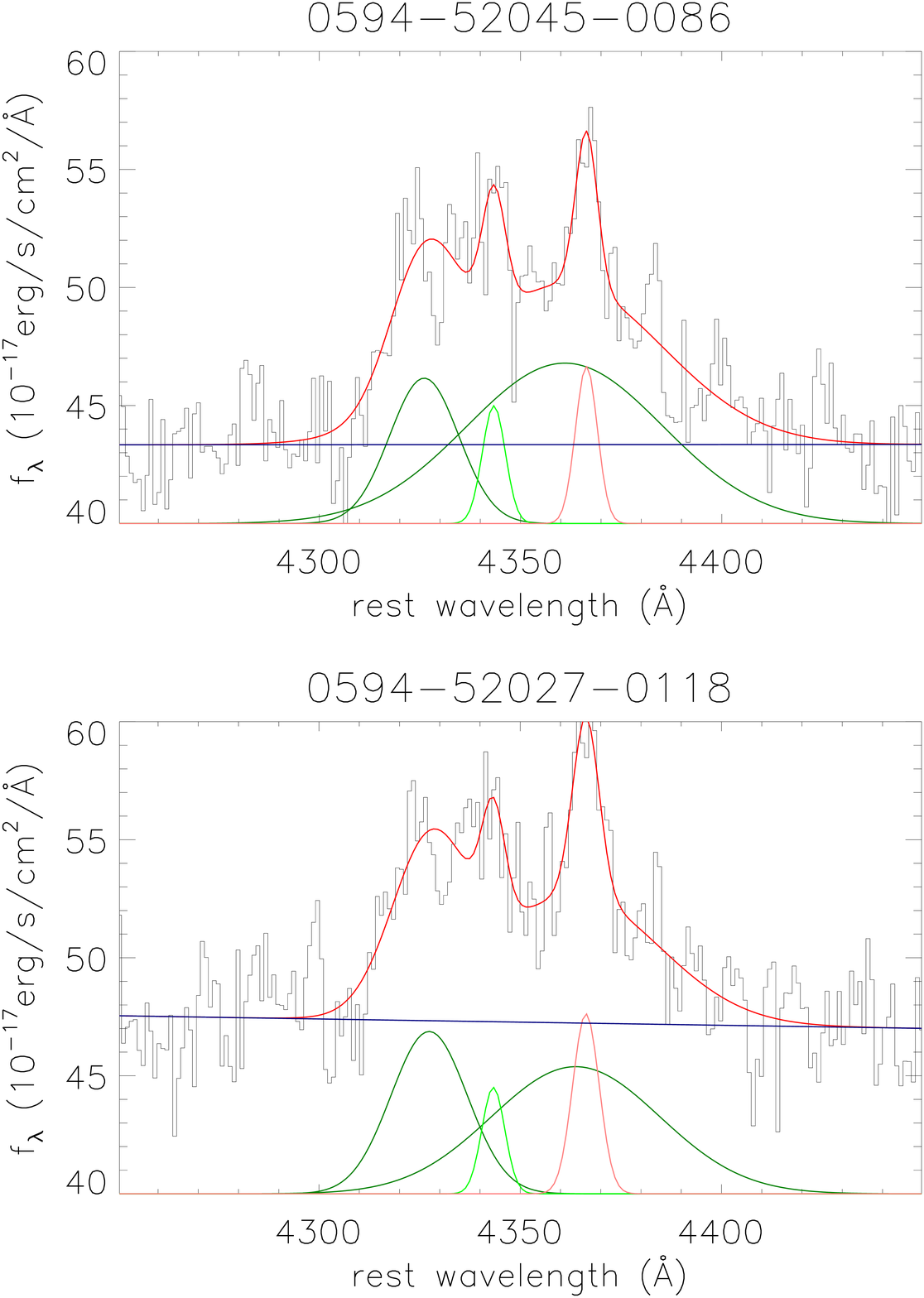}
\caption{Best descriptions to the emission lines around H$\gamma$ in MJD=52045 (top panel) and in 
MJD=52027 (bottom panel). In each panel, solid black line and solid red line show the line spectrum 
and the determined best descriptions, respectively. Solid dark blue line shows the determined AGN 
continuum emissions, solid dark green lines show the determined two broad Gaussian components in 
the broad H$\gamma$, solid green line shows the determined narrow H$\gamma$, solid pink lines show 
the determined [O~{\sc iii}]$\lambda4364$\AA.
}
\label{hg}
\end{figure}

\begin{figure}
\centering\includegraphics[width = 8cm,height=10cm]{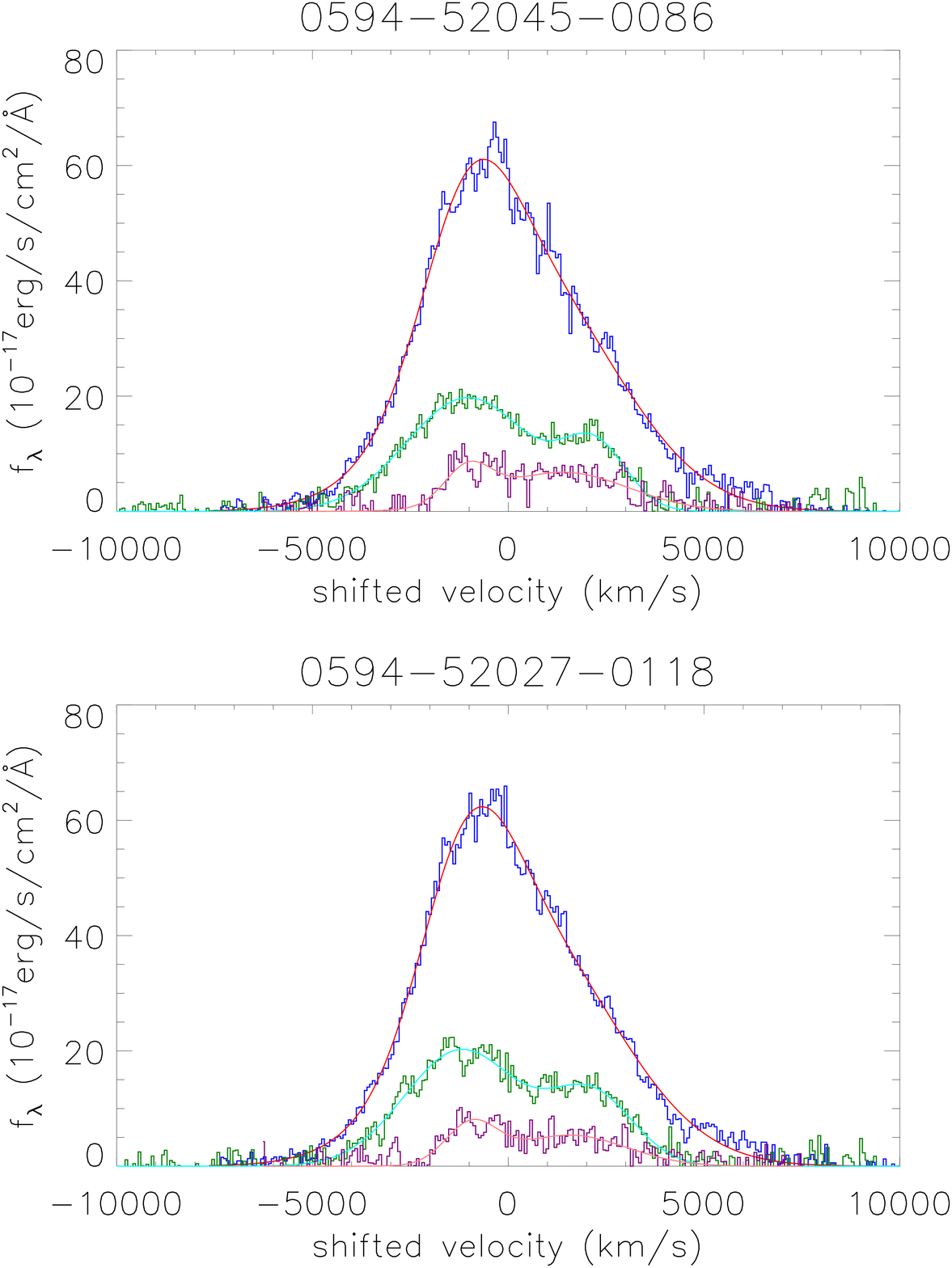}
\caption{Line profiles of the broad H$\alpha$, the broad H$\beta$ and the broad H$\gamma$ in 
velocity space in MJD=52045 (top panel) and in MJD=52027 (bottom panel), after subtractions of 
both the narrow emission line components and the continuum emissions component. In each panel, 
solid line in blue, dark green and purple shows the line profile of the broad H$\alpha$, of the 
broad H$\beta$ and of the broad H$\gamma$, respectively, solid line in red, cyan and pink shows 
the corresponding best-fitting results to the broad H$\alpha$, the broad H$\beta$ and the broad 
H$\gamma$ by the two broad Gaussian components shown as solid dark green lines in Figure~\ref{hb}, 
in Figure~\ref{ha} and in Figure~\ref{hg}, respectively. 
}
\label{hab}
\end{figure}

\section{Spectroscopic Properties in SDSS J1547} 

    SDSS J1547 at redshift of 0.09768 has been observed twice in one month in SDSS \citep{gun06, aa20}. 
The SDSS spectra (information of PLATE-MJD-FIBERID: 0594-52027-0118 and 0594-52045-0086) are collected from SDSS 
DR16 (Data Release 16) and shown in Figure~\ref{spec}. Based on the spectroscopic features, SDSS 
J1547 is a broad line AGN with apparently broad Balmer emissions. And through the shown spectra, 
there is an interesting point that there is an additional peak (marked by vertical dashed red 
line in Figure~\ref{spec}) in the broad H$\beta$ but not in the broad H$\alpha$. The double-peaked 
broad H$\beta$ but single-peaked broad H$\alpha$ can be confirmed in the repeated spectra of SDSS 
J1547, indicating the strange features could be not due to observational mistakes. Meanwhile, 
SDSS J1547 (also named as MS 1545.3+0305) has also been observed by the Clay Telescope and shown 
in Figure 1\ in \citet{hk09} with the apparent double-peaked features in the broad H$\beta$.

    In order to well confirm the different line profiles of the broad H$\beta$ and the broad 
H$\alpha$, the following model functions are accepted to describe the emission lines around 
H$\beta$ and around H$\alpha$. Similar as what we have done in the more recent paper \citet{zh21}, 
for the emission lines around H$\beta$ with rest wavelength range from 4400\AA\ to 5600\AA, there 
are two broad Gaussian functions applied to describe the broad H$\beta$, three narrow Gaussian 
functions applied to describe the narrow H$\beta$ and the core \oiii components, two another 
Gaussian functions applied to describe the broad \oiii components, two another additional broad 
Gaussian functions applied to describe the extremely extended broad \oiii components (which will 
be discussed in detail in the follows), one broad Gaussian function applied to describe weak 
He~{\sc ii} line, one power law function $P_\lambda=\alpha\times(\lambda/5100\text{\AA})^\beta$ 
applied to describe AGN continuum emissions, and the broadened and scaled Fe~{\sc ii} templates 
discussed in \citet{kp10} applied to describe the strong optical Fe~{\sc ii} lines. Based on 
the widely applied Levenberg-Marquardt least-squares minimization technique, the best-fitting 
results to the emission lines around H$\beta$ can be well determined, and shown in Figure~\ref{hb}, 
with the determined $\chi^2=SSR/Dof$ (where $SSR$ and $Dof$ as summed squared residuals and 
degree of freedom) about 1.41 and 1.06 for the emission lines observed in MJD=52045 and in 
MJD=52027, respectively. The determined line parameters of central wavelength, second moment 
and line flux of each emission component are listed in Table~1.

    The emission lines around H$\alpha$ with rest wavelength range from 6400\AA~ to 6800\AA~ 
can be well described by the following model functions. There are two  broad Gaussian functions 
applied to describe the broad H$\alpha$, five narrow Gaussian functions applied to describe 
the narrow H$\alpha$, the [N~{\sc ii}] doublet and the [S~{\sc ii}] doublet. Then, through 
the Levenberg-Marquardt least-squares minimization technique, the best descriptions to the 
emission lines around H$\alpha$ can be well determined and shown in Figure~\ref{ha}, with 
the determined $\chi^2=SSR/Dof$ about 1.81 and 0.99 for the emission lines observed in MJD=52045 
and in MJD=52027, respectively. The corresponding line parameters are also listed in Table~1.

    Before proceeding further, we show further discussions on the extremely extended broad 
\oiii components around 4975\AA~ shown as solid purple lines in Figure~\ref{hb}. If the broad 
components were not the \oiii emission components, but from broad H$\beta$ emissions, it 
could be expect that there should be similar broad components in the broad H$\alpha$ around 
6720\AA. However, such expected broad emission components in broad H$\alpha$ cannot be found. 
Therefore, the broad components around 4975\AA~ are accepted as the extremely extended broad 
\oiii components in SDSS J1547, not part of the broad H$\beta$ emissions.

   Meanwhile, properties of broad H$\gamma$ are also considered. The best-fitting results to 
the emission lines around H$\gamma$ are shown in Figure~\ref{hg}, based on the similar model 
functions: two broad Gaussian functions for the broad H$\gamma$, one narrow Gaussian function 
for the narrow H$\gamma$ and one narrow Gaussian function for the [O~{\sc iii}]$\lambda4364$\AA. 
The determined line parameters are also listed in Table~1. Then, in order to clearly show the 
compared line profiles between the broad Balmer emission lines, the line profiles are shown 
in velocity space in Figure~\ref{hab}. It is obvious that there are two peaks around -1103${\rm km/s}$ 
and 2180${\rm km/s}$ in the broad H$\beta$ and also in the broad H$\gamma$, but only one 
apparent peak around -750${\rm km/s}$ in the broad H$\alpha$.

	Whether the different line profiles of broad Balmer emission lines could be due to effects of 
different optical depths of broad Balmer emission lines? The effects of different optical depths 
have been well discussed in \citet{kg04, ben10} and in more recent discussed results in \citet{nh20}, 
leading to expected longer distance of emission regions and smaller line width of broad H$\alpha$ than 
those of broad H$\beta$ and broad H$\gamma$, because broad H$\alpha$ has the largest optical depth among the 
broad Balmer emission lines. Therefore, the line widths of broad H$\alpha$, broad H$\beta$ and broad 
H$\gamma$ are firstly checked. Based on the shown line profiles $p_\lambda$ of the broad Balmer 
emission lines in Figure~\ref{hab} and the definition of second moment $\sigma_l$ of emission lines in 
\citet{pe04},
\begin{equation}
\begin{split}
&\lambda_0~=~\frac{\int~\lambda~p_\lambda d\lambda}{\int~p_\lambda d\lambda} \\
&\sigma_l^2~=~\frac{\int~\lambda^2~p_\lambda d\lambda}{\int~p_\lambda d\lambda}~-~\lambda_0^2
\end{split}
\end{equation}, 
the line widths are measured as 47.8\AA~ (2184${\rm km/s}$), 31.9\AA~ (1968${\rm km/s}$) and 
26.3\AA~ (1817${\rm km/s}$) of broad H$\alpha$, broad H$\beta$ and broad H$\gamma$, respectively. 
It is clear that broad H$\alpha$ has line width (second moment) larger than broad H$\beta$ and broad 
H$\gamma$, not consistent with the expected results after considerations of effects of different 
optical depths of broad Balmer emission lines. Meanwhile, the FWHMs (full widths at half maximum) 
of the broad Balmer emission lines can be measured as 101.7\AA~ (4648${\rm km/s}$), 90.9\AA~ 
(5608${\rm km/s}$) and 67.8\AA~ (4685${\rm km/s}$) of broad H$\alpha$, broad H$\beta$ and broad 
H$\gamma$, respectively. Even considering the FWHMs as the line width, although the broad H$\alpha$ 
is broader than the broad H$\beta$, the broad H$\beta$ quite narrower than the broad H$\gamma$, 
not consistent with the expected results after considerations of effects of different optical 
depths of broad Balmer emission lines. Therefore, effects of different optical depths of broad 
Balmer emission lines can not be preferred to explain the different profiles of broad Balmer emission lines, 
and there are no further discussions on the effects in the manuscript.

    Based on the results shown in Figure~\ref{hb}, Figure~\ref{ha} and Figure~\ref{hg} and in 
Figure~\ref{hab}, the very different line profiles can be well detected between the broad 
H$\alpha$ and the broad H$\beta$. Furthermore, based on the measured line parameters, the flux 
ratio of total broad H$\alpha$ to total broad H$\beta$ is about 4.08, and the flux ratio of 
narrow H$\alpha$ to narrow H$\beta$ is about 5.7. The larger flux ratio in narrow Balmer lines 
than in broad Balmer lines indicates that the flux ratio of broad Balmer lines is not due to 
commonly accepted obscurations, otherwise there should be larger flux ratio of broad H$\alpha$ 
to broad H$\beta$ than the ratio of narrow H$\alpha$ to narrow H$\beta$. Furthermore, in spite 
of the lower spectral quality around H$\gamma$, the larger flux ratio 14 of the narrow H$\alpha$ 
to the narrow H$\gamma$ than the flux ratio 12 of the broad H$\alpha$ to the broad H$\gamma$, 
providing further evidence that the commonly accepted obscurations should be not preferred 
in \obj. Therefore, accretion disk origin and different intrinsic physical conditions are 
mainly considered to explain the different line profiles between the broad Balmer lines in \obj.

\begin{table}
\caption{Line parameters of emission lines of \obj}
\begin{tabular}{lccc}
\hline\hline
\multicolumn{4}{c}{parameters in MJD=52045}  \\
\hline
Line    &   $\lambda_0$  &   $\sigma$  & flux  \\
\hline
H$\alpha_{B1}$ & 6545.7$\pm$0.9 & 25.8$\pm$1.1 & 1683$\pm$151 \\
H$\alpha_{B2}$ & 6576.9$\pm$1.2  & 50.9$\pm$0.6 &  5118$\pm$159   \\	
H$\beta_{B1}$ &  4848.4$\pm$0.9 & 25.6$\pm$0.9 & 1267$\pm$46 \\
H$\beta_{B2}$ &  4900.7$\pm$0.9 & 13.2$\pm$1.1 & 357$\pm$44 \\
H$\gamma_{B1}$ &  4326.1$\pm$1.1 & 8.6$\pm$1.6 & 132$\pm$42 \\
H$\gamma_{B2}$ &  4361.1$\pm$3.9 & 24.2$\pm$3.9 & 415$\pm$72 \\
\hline
H$\alpha_{N}$ & 6567.9$\pm$0.1 & 2.6$\pm$0.1 & 487$\pm$15 \\
H$\beta_{N}$ & 4864.9$\pm$0.1  & 1.6$\pm$0.1 & 77$\pm$7 \\
H$\gamma_{N}$ & 4343.4$\pm$0.4  & 2.8$\pm$0.8 & 36$\pm$12 \\
\o34 & 4366.4$\pm$0.4 & 2.7$\pm$0.5 & 45$\pm$9 \\
\oc  &  5010.5$\pm$0.1 & 2.3$\pm$0.1 & 463$\pm$39 \\ 
\ob & 5006.9$\pm$0.3 & 4.8$\pm$0.2 & 536$\pm$43 \\
\oe & 4970.8$\pm$3.3 & 37.5$\pm$3.3 & 542$\pm$41 \\
\nii  &  6588.6$\pm$0.1 & 2.9$\pm$0.1 & 444$\pm$14 \\ 
\sii  & 6721.6$\pm$0.1 & 3.3$\pm$0.1 & 181$\pm$7 \\ 
\si  & 6736.3$\pm$0.2 & 3.3$\pm$0.1 & 145$\pm$6 \\ 
\hline\hline
\multicolumn{4}{c}{parameters in MJD=52027}  \\
\hline\hline
H$\alpha_{B1}$ & 6545.2$\pm$1.1 & 25.8$\pm$1.2 & 1830$\pm$194 \\
H$\alpha_{B2}$ & 6577.4$\pm$1.6 & 50.7$\pm$0.8  & 5049$\pm$206 \\
H$\beta_{B1}$ & 4845.6$\pm$1.6 & 23.5$\pm$1.3  & 1188$\pm$74 \\
H$\beta_{B2}$ & 4899.4$\pm$1.8 & 17.3$\pm$1.6  & 537$\pm$78 \\
H$\gamma_{B1}$ & 4327.4$\pm$1.6 & 9.5$\pm$1.8  & 163$\pm$49 \\
H$\gamma_{B2}$ & 4363.7$\pm$4.8 & 20.9$\pm$5.1  & 282$\pm$66 \\
\hline
H$\alpha_{N}$ & 6568.1$\pm$0.1  & 2.6$\pm$0.1 & 470$\pm$18 \\
H$\beta_{N}$ & 4865.1$\pm$0.2  & 2.2$\pm$0.2 &  93$\pm$11 \\
H$\gamma_{N}$ & 4343.3$\pm$0.5  & 2.8$\pm$0.8 &  32$\pm$11 \\
\o34 & 4366.3$\pm$0.5 &  3.3$\pm$0.7 &  63$\pm$17 \\
\oc  &  5010.7$\pm$0.1 & 2.4$\pm$0.2 & 502$\pm$75 \\
\ob & 5006.7$\pm$0.7 & 4.5$\pm$0.3 & 481$\pm$80 \\
\oe & 4980.6$\pm$4.8 & 36.1$\pm$4.2 & 495$\pm$49 \\
\nii  &  6588.8$\pm$0.1 & 2.9$\pm$0.1 & 438$\pm$16 \\
\sii  & 6722.3$\pm$0.2  & 3.2$\pm$0.1 & 174$\pm$8 \\
\si  & 6736.4$\pm$0.2 & 3.2$\pm$0.1 & 147$\pm$8 \\
\hline
\end{tabular}\\
{\bf Note:} The second, third and fourth columns show the rest central wavelength in unit 
of \AA, the line width (second moment) in unit of \AA\ and the line flux in unit of 
$10^{-17}{\rm erg/s/cm^2}$ of the emission line components. \\
The suffix "B1" and "B2" represent the two broad Gaussian components in the broad Balmer 
lines. The suffix "N" represents the narrow Gaussian component in the narrow Balmer lines. 
The \oc, \ob~ and \oe~ represent the determined core, broad and the extremely extended 
broad \oiii components. 
\end{table}

\section{Accretion disk origin of the double-peaked broad Balmer emission lines of \obj?}

   Based on the results above, it can be confirmed that broad H$\beta$ has double-peaked emission 
features. Therefore, in the section, it is interesting to check whether the commonly accepted accretion 
disk origin can be applied to explain the line profiles of broad Balmer emission lines in \obj. 

\begin{figure*}
\centering\includegraphics[width = 18cm,height=6cm]{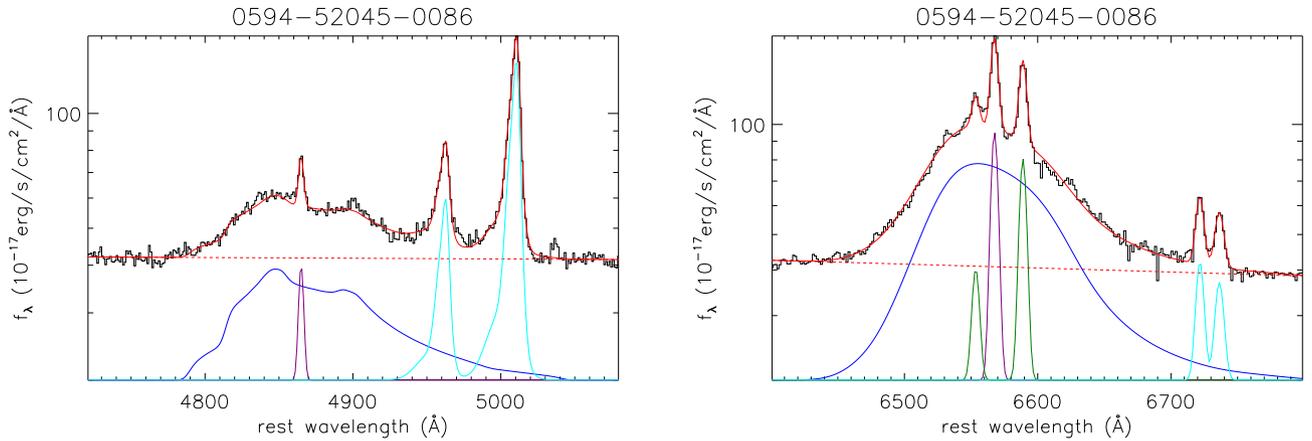}
\caption{The accretion disk model determined best descriptions to the double-peaked 
broad H$\beta$ in the left panel and to the broad H$\alpha$ in the right panel. In the left 
panel, solid black line shows the line spectrum after subtractions of the optical Fe~{\sc ii} lines. 
In the right panel, solid black line shows the line spectrum. In each panel, solid red line shows 
the best descriptions to the emission lines by the elliptical accretion disk model plus multiple 
Gaussian functions, dashed red line shows the determined AGN continuum emissions underneath the 
emission lines, solid blue line shows the determined broad component from the disk-like emission 
regions, solid purple line shows the determined narrow Balmer line. In the left panel, solid cyan lines 
show the determined [O~{\sc iii}] doublet. In the right panel, solid lines in dark green and in cyan
shows the determined [N~{\sc ii}] and [S~{\sc ii}] doublets. {\bf In order to show clearly determined 
broad Balmer emission lines, the Y-axis is in logarithmic coordinate.}
}
\label{disk}
\end{figure*}

	In order to explain the double-peaked broad emission lines of AGN, besides the well described 
BBH model in the Introduction and well discussed in the following section, the proposed accretion disk 
origin has been well accepted. The accretion disk model is firstly proposed in \citet{ch89, chf89} to 
explain the double-peaked broad emission lines in AGN Arp102B and well applied in \citet{eh94}. Then, 
different relativistic accretion disk models have been proposed in the literature, such as the improved 
elliptical accretion disk model in \citet{el95}, the circular disk model plus contributions of spiral arms 
in \citet{st03}, the disk model with considerations of warped structures in \citet{hb00}, the stochastically 
perturbed accretion disk model in \citet{fe08}, etc. Here, the elliptical accretion disk model (without 
contributions of subtle structures) well discussed in \citet{el95} is preferred, because the model 
can be applied to well explain the double-peaked broad H$\beta$ in \obj. There are seven model parameters 
in the elliptical accretion disk model, inner boundary $r_0$ and out boundary $r_1$ in the units of 
$R_g$ (Schwarzschild radius), inclination angle $i$ of disk-like BLRs, eccentricity $e$, orientation 
angle $\phi_0$ of elliptical rings, local broadening velocity $\sigma_L$, line emissivity slope $q$ 
($f_r~\propto~r^{-q}$). Meanwhile, we have also applied the very familiar elliptical accretion disk model, 
see our studies on double-peaked emission lines in \citet{zh05, zh11, zh13, zh13a, zh15, zh21a}. More 
detailed descriptions on the applied elliptical accretion disk model can be found in \citet{el95, st03, sv03}, 
and there are no further descriptions on the elliptical accretion disk model in the manuscript. And 
the double-peaked broad H$\beta$ is described by the elliptical accretion disk model as follows.

    Through the Levenberg-Marquardt least-squares minimization method, the emission lines can be 
described by the elliptical accretion disk model for the broad H$\beta$ plus multiple Gaussian 
functions for the narrow lines of narrow H$\beta$, core, broad and extremely extended components 
in the \oiii doublets. When the elliptical accretion disk model is applied, the seven model parameters 
have restrictions as follows. The inner inner boundary $r_0$ is larger than 15${\rm R_G}$ and 
smaller than 1000${\rm R_G}$. The out boundary $r_1$ is larger than $r_0$ and smaller than 
$10^6{\rm R_G}$. The inclination angle $i$ of disk-like emission regions of broad H$\beta$ 
has $\sin(i)$ larger than 0.05 and smaller than 0.95. The eccentricity $e$ is larger than 0 and 
smaller than 1. The orientation angle $\phi_0$ of elliptical rings is larger than 0 and smaller 
than $2\pi$. The local broadening velocity $\sigma_L$ is larger than 10${\rm km/s}$ and smaller 
than $10^4{\rm km/s}$. And the line emissivity slope $q$ ($f_r~\propto~r^{-q}$) is larger than -7 
and smaller than 7. The model parameters of the disk-like emission regions are about 
$r_0=163\pm10{\rm R_g}$, $r_1=3265\pm391{\rm R_g}$, $\sin(i)=0.293\pm0.004$, 
$\sigma_L=175\pm40{\rm km/s}$, $q=2.31\pm0.06$, $e=0.621\pm0.026$, $\phi_0=3.19$, respectively. 
The best descriptions to the double-peaked broad H$\beta$ is shown in the left panel of 
Figure~\ref{disk} with $\chi^2\sim1.01$.

	Now considering probably different emission regions of broad H$\alpha$ from broad H$\beta$, 
it is interesting to check whether similar disk-like emission regions can be applied to explain 
the observed line profile of broad H$\alpha$. Here, the similar accretion disk model plus multiple 
Gaussian functions are applied, but the accretion disk model parameters of $r_0=163$, 
$\sin(i)=0.293$, $e=0.621$ and $\phi_0=3.19$ being fixed. The best descriptions to the 
broad H$\alpha$ is shown in the right panel of Figure~\ref{disk} with $\chi^2\sim1.91$. And the 
other three determined model parameters of the disk-like emission regions are about 
$r_1=2872\pm250{\rm R_g}$, $\sigma_L=700\pm50{\rm km/s}$, $q=1.72\pm0.06$, respectively.

     The accretion disk model can be well applied to described the different line profiles 
of broad Balmer emission lines, but the similar disk-like emission regions (similar boundaries, 
similar inclinations, similar eccentricity, etc.) for the broad H$\alpha$ and the broad H$\beta$ 
have much different local broadening velocities, which cannot be naturally expected. In one word, 
in order to overwhelm the expected double-peaked features in broad H$\alpha$, quite large local 
broadening velocity should be necessary in the emission regions of broad H$\alpha$. Unless there 
are quite different emission regions for the broad H$\alpha$ and the broad H$\beta$, the quite 
different broadening velocities can be not expected in emission regions of broad H$\alpha$ and 
broad H$\beta$. Therefore, the accretion disk origin is not preferred to explain the different 
line profiles of broad Balmer emission lines in \obj.

\section{Two BLRs related to a central BBH system}

   Physical conditions have strong effects on intrinsic flux ratio of broad H$\alpha$ to broad 
H$\beta$, such as the previous results in \citet{rf89, kg04, nh20}. In order to provide different 
physical conditions to broad Balmer line emission regions, two BLRs around central two black holes 
in a central BBH system should be well preferred. Intrinsic flux ratio of broad H$\alpha$ to broad 
H$\beta$ ($f_{\rm ab}$) in different physical conditions can be well varied from around 2 to 
around 4, such as the more recent discussed results in \citet{nh20}. Now it is interesting to 
check whether varying $f_{\rm ab}$ from around 2 to around 4 can lead to different line profiles 
between Broad Balmer lines under the assumption of a central BBH system in \obj.

\begin{figure*}
\centering\includegraphics[width = 18cm,height=5cm]{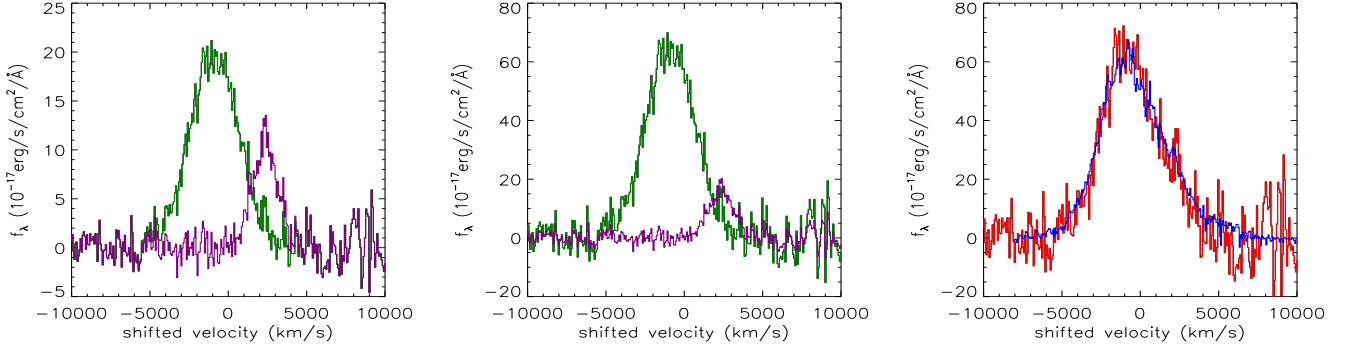}
\caption{Left panel shows the two broad components in the broad H$\beta$: the blue-shifted 
broad component $F_{\rm Hb1}$ shown as solid dark green line and the red-shifted broad component 
$F_{\rm Hb2}$ shown as solid purple line. Middle panel shows the scaled broad components of 
$F_{\rm Hb1}\times3.3$ and $F_{\rm Hb2}\times1.5$. Right panel shows the expected line profile 
of the broad H$\alpha$ by $F_{\rm Hb1}\times3.3+F_{\rm Hb2}\times1.5$ shown as solid red line. 
And in right panel, solid blue line shows the observed line profile of the broad H$\alpha$ in \obj.
}
\label{exp}
\end{figure*}

\begin{figure}
\centering\includegraphics[width = 8cm,height=15cm]{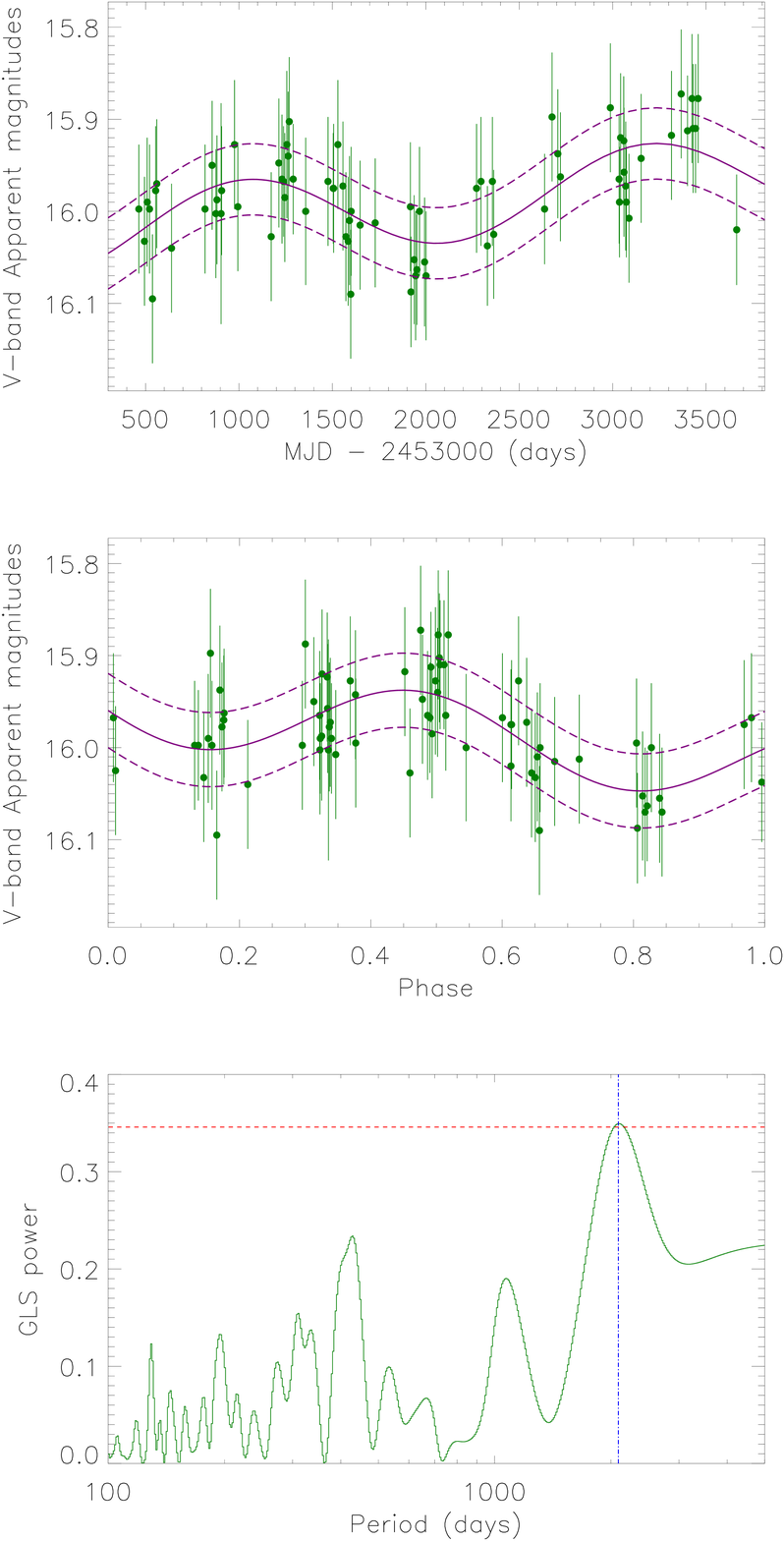}
\caption{Top panel shows the long-term light curves from CSS (in dark green). Middle panel 
shows the corresponding phase folded light curve based on the determined periodicity about 
2159days. In top and middle panels, solid and dashed lines in purple show the best descriptions 
to the light curve and the corresponding $1\sigma$ confidence bands, based on a sinusoidal 
function plus a linear trend. Bottom panel shows the Generalized Lomb-Scargle (GLS) periodogram. 
In bottom panel, horizontal dashed red line shows the 99.99\% confidence level (0.0001 as 
the false-alarm probability), vertical blue line marks the peak around 2091days by the periodogram.
}
\label{lmc}
\end{figure}

   Not considering complicated structures, the two components are accepted from the broad H$\beta$: 
the blue-shifted broad component $F_{\rm Hb1}$ and the red-shifted broad component $F_{\rm Hb2}$,
then, it is interesting to check whether two different flux ratios $f_{\rm ab}$ on $F_{\rm Hb1}$ and 
$F_{\rm Hb2}$ can lead to the similar line profile of the broad H$\alpha$. We can simply find that 
$f_{\rm ab}\sim3.3$ on the blue-shifted broad component and $f_{\rm ab}\sim1.5$ on the red-shifted 
broad component can lead to the similar line profile of the observed broad H$\alpha$. The results 
are shown in Figure~\ref{exp}. Here, $F_{\rm Hb1}$ and $F_{\rm Hb2}$ are similar as the smooth 
Gaussian components shown as solid dark green lines in Figure~\ref{hb}, but by the following formula
\begin{equation}
\begin{split}
F_{\rm Hb1} &= F_{\lambda}- F_{\rm B2} - F_{\rm N} - pow \\ 
F_{\rm Hb2} &= F_{\lambda}- F_{\rm B1} - F_{\rm N} - pow
\end{split}
\end{equation}
where $F_{\lambda}$ means the line spectrum shown in Figure~\ref{hb}, $F_{\rm B1}$ means the 
determined blue-shifted broad Gaussian component in the broad H$\beta$, $F_{\rm B2}$ means the 
determined red-shifted broad Gaussian component in the broad H$\beta$, $F_{\rm N}$ means the 
sum of the determined narrow emission line components and $pow$ means the determined power-law 
component. Therefore, two BLRs with different physical conditions can be well applied to explain 
the observed different line profiles between the broad H$\alpha$ and the broad H$\beta$. Here, 
the different physical conditions include different ionization parameters, different electron 
temperatures, different electron densities, etc.. In the current stage, it is hard to determine 
which parameter has the key role on the different $f_{\rm ab}$ to the two BLRs in the central 
BBH system.

   If the different line profiles of the broad Balmer emission lines were related to a central 
BBH system, the interesting features on the different line profiles could be detected in the 
candidates for BBH systems. Actually, there is really one BBH system reported with different 
line profiles of the broad Balmer emission lines in SDSS J0159+0105 as simply mentioned in 
\citet{zb16}. In SDSS J0159+0105, the red bump in the broad H$\beta$ is significant, but it 
is not very significant in the broad H$\alpha$, similar as the case in \obj.

   Once accepted the central BBH system in \obj, it is interesting to check whether expected 
optical quasi-periodic oscillations (QPOs) can be detected in the long-term variabilities in 
\obj. The long-term V-band light curve has been collected from Catalina Sky Survey (CSS) 
\citep{dr09} \url{http://nesssi.cacr.caltech.edu/DataRelease/} with MJD-2453000 from 464 
(April 2005) to 3644 (January 2014), and shown in top panel of Figure~\ref{lmc}. Through the 
Levenberg-Marquardt least-squares minimization technique, the long-term CSS V-band light curve 
can be well described by a sinusoidal function plus a linear trend,
\begin{equation}
\begin{split}
LMC&=(16.028\pm0.019)-(0.018\pm0.009)\times\frac{t}{\rm 1000days} \\
	&+ (0.044\pm0.012)\sin(\frac{2\pi t}{2159\pm210}+(1.725\pm0.561))
\end{split}
\end{equation}
leading to the QPOs with periodicity about $2159\pm210$ days. Meanwhile, based on the 
determined periodicity, the phase folded light curve is shown in the bottom panel of 
Figure~\ref{lmc}, which can also be well described by a sinusoidal function. The results 
on the directed fitted results by the sinusoidal function and the phase folded light curve 
well described by a sinusoidal function strongly support the optical QPOs in \obj. However, 
the time duration of the CSS light curve is only 1.5times longer than the determined 
periodicity, leading the determined QPOs not to have high confidence levels. Future 
monitoring on \obj will be necessary to check the expected QPOs. Therefore, the 
Generalized Lomb-Scargle periodogram \citep{ln76, sj82, zk09, zb16} is applied and shown 
in bottom panel of Figure~\ref{lmc} with the clear peak around 2091days with confidence 
level higher than 99.99\% well consistent with the determined $2159\pm210$ days by the 
direct-fitting procedure, but no further discussions on the results from the periodogram.

    Based on the different line profiles of the broad Balmer emission lines and the 
expected long-term optical QPOs, the BBH system can be well preferred as the first choice 
in \obj. Then, under the assumption of a BBH system, further properties of the BBH system 
can be simply discussed as follows. First, under the Virialization assumptions to the 
broad Balmer line emission clouds of each central BLRs \citep{pe04, gh05, vp06} combining 
with the more recent R-L relation discussed in \citet{ben13}:
\begin{equation}
\frac{M_{\rm BH}}{\rm M_\odot}=3.6\times10^6\times(\frac{L_{\rm H\beta}}{\rm 10^{42}erg/s})^{0.56}
    \times(\frac{FWHM_{\rm H\alpha}}{\rm 1000km/s})^{2}
\end{equation},
based on the two apparent broad components in the broad H$\beta$, the virial BH masses of 
the two central BHs can be estimated as $M_{\rm BH,1}\sim2.3\times10^7{\rm M_\odot}$ and 
$M_{\rm BH,2}\sim5.1\times10^6{\rm M_\odot}$ for the blue-shifted BH accreting system and 
for the red-shifted BH accreting system, respectively. Second, based on the 
expected periodicity about 2159days combining with the estimated BH masses of the central 
BHs in a BBH system, the space separation of the two central BHs $A_{BBH}$ can be estimated 
as 
\begin{equation}
A_{BBH}=0.432\times M_{8}\times(\frac{P_{BBH}/year}{2652M_{8}})^{2/3}\sim0.005pc
\end{equation}
where $M_{8}$ represents the total BH mass of the BBH system in unit of $10^8{\rm M_\odot}$ 
and $P_{BBH}\sim5.92{\rm years}$ represents the orbital period of the BBH system in \obj. 
Third, based on the R-L relation in \citet{ben13}, expected BLRs size could be 
$R_{BLRs}\sim25{\rm light-days}$ through the continuum luminosity at 5100\AA~ about 
$5.04\times10^{43}{erg/s}$ in \obj. The very larger $R_{BLRs}\sim25{\rm light-days}$ 
than the estimated $A_{BBH}\sim5.8{\rm light-days}$ could probably rule out the existence 
of two totally distinctive BLRs in the central BBH system in \obj.

   Before proceeding further, we simply discuss the larger $R_{BLRs}\sim25{\rm light-days}$ 
than the estimated $A_{BBH}\sim5.8{\rm light-days}$ in \obj~ as follows. As discussed results 
in \citet{sl10}, line profiles of broad emissions in BBHs systems are more complex, when the 
central two BLRs can no longer be distinct. However, in \obj, broad H$\alpha$ and broad 
H$\beta$ can be well described by two concise Gaussian components, providing weak signs of mixed two 
BLRs. The results probably indicate either $R_{BLRs}$ or $A_{BBH}$ is not so reliable in \obj. 
The R-L empirical relation determined from normal broad line AGN could not be well applied in BBHs 
systems. In the near future, further measurements on time lags between variabilities of broad 
Balmer emission lines and continuum emissions could provide more clearer results on sizes of 
broad Balmer line emission regions. Moreover, more accurate virial BH mass could also be 
estimated through multiple spectroscopic results. More efforts are necessary to
determine more clearer properties of the expected BBH system in \obj.

\section{How many AGN similar as \obj~ can be found?}

    Before the end of the manuscript, it is interesting to consider the following question 
what percentage of broad line AGN can be expected to have very different line profiles 
of broad Balmer emission lines as indicators to central BBH system. Until now, \obj~ is 
the only one object with reported detailed discussed different Balmer emission line 
profiles related to a central BBH. In the near future, it is necessary and interesting 
to detect more objects with very different line profiles of broad Balmer emission lines, 
and to check whether the objects could harbour central BBH systems. Here, the expected 
percentage can be estimated as follows. Once assumed central BBHs, the simple two broad 
Gaussian described components are accepted in the broad H$\beta$. Then, series of 20000 
fake broad Balmer lines can be created by two steps. The blue-shifted broad Gaussian 
component $F_{H\beta, b}$ in the broad H$\beta$ has parameters of central wavelength 
randomly from -3000${\rm km/s}$ to 0${\rm km/s}$ in velocity space, second moment randomly 
from 800${\rm km/s}$ to 3000${\rm km/s}$ and line flux as 1\ in arbitrary unit. And the 
red-shifted broad Gaussian component $F_{H\beta, r}$ in the broad H$\beta$ has parameters 
of central wavelength randomly from 0${\rm km/s}$ to 3000${\rm km/s}$ in velocity space, 
second moment randomly from 800${\rm km/s}$ to 3000${\rm km/s}$ and line flux randomly 
from 0.25 to 4\ in arbitrary unit. The fake broad H$\beta$ can be created as 
$F_{H\beta, b} + F_{H\beta, r}$. Then, the fake two broad broad Gaussian components 
$F_{H\alpha, b}$ and $F_{H\alpha, r}$ in the velocity space are determined by 
$F_{H\alpha, b}=F_{H\beta, b}\times f_{\rm ab, b}$ and 
$F_{H\alpha, r}=F_{H\beta, r}\times f_{\rm ab, r}$ with $f_{\rm ab, b}$ and $f_{\rm ab, r}$ 
randomly from 1.5 to 4, leading the corresponding fake broad H$\alpha$ to be 
$F_{H\alpha, b} + F_{H\alpha, r}$.

Then, based on the fake broad Balmer lines, the line profiles of the fake broad Balmer 
lines can be well checked by properties of the three parameters: the absolute difference 
of central wavelength $\Delta\lambda_0$ between the broad H$\alpha$ and the broad H$\beta$, 
the absolute difference of second moment $\Delta\sigma$ between the broad H$\alpha$ and 
the broad H$\beta$, and the number of peaks $N_{\beta}$ and $N_{\alpha}$ in the broad 
H$\beta$ and the broad H$\alpha$. For the case in \obj, the determined values are 
$\Delta\lambda_{0,*}\sim387{\rm km/s}$, $\Delta\sigma_*\sim215{\rm km/s}$, $N_{\beta,*}=2$ 
and $N_{\alpha,*}=1$. Then, based on the criteria that $\Delta\lambda_0\ge\Delta\lambda_{0,*}$, 
$\Delta\sigma\ge\Delta\sigma_*$, $N_{\beta}=N_{\beta,*}$ and $N_{\alpha}=N_{\alpha,*}$, 
there are 25 cases with double-peaked broad H$\beta$ but single-peaked broad H$\alpha$, 
among the 20000 fake broad Balmer emission lines, indicating about 0.125\% ($25/20000$) 
of broad line AGN having very different broad Balmer emission lines, similar as the case 
in \obj. Therefore, among the 13000 quasars in SDSS DR16 with redshift less than 0.35, 
there could be at least 16 quasars with very different broad Balmer emission lines. It 
will be worth to detect and check the SDSS quasars with very different line profiles of 
broad Balmer lines under the assumptions of central BBH systems in the near future.

\section{Conclusions}

   Finally, we give our main conclusions as follows. 
\begin{itemize}   
\item Through the high quality SDSS spectra, very different broad Balmer emission lines 
can be confirmed in \obj: double-peaked broad H$\beta$ and broad H$\gamma$ but single-peaked 
broad H$\alpha$. 
\item The determined flux ratio of the narrow H$\alpha$ to the narrow H$\beta$ is larger 
than the ratio of the broad H$\alpha$ to the broad H$\beta$, indicating that rather than 
effects of dust obscurations, two BLRs related to a central BBH system are preferred in \obj. 
And the different physical conditions on the two expected central BLRs related to a central 
BBH system can be well applied to describe the very different broad Balmer emission lines 
in \obj.
\item The long-term CSS V-band light curve of \obj~ is checked. The light curve and the 
corresponding phase-folded light curve can be well described by a sinusoidal function, 
indicating probable QPOs with periodicity about $2159\pm210$days expected by a central 
BBH system. 
\item The expected BBH system can be estimated with virial BH masses about 
$2.3\times10^7{\rm M_\odot}$ and $5.1\times10^6{\rm M_\odot}$ and with space separation 
about 0.005pc. 
\item Based on randomly created fake broad Balmer emission lines, about 0.125\% of broad 
line AGN (quasars) have very different broad Balmer emission lines, similar as the case in \obj. 
\end{itemize}

\section*{Acknowledgements}
Zhang gratefully acknowledges the anonymous referee for carefully reading our 
manuscript with patience, and giving us constructive comments and suggestions to greatly 
improve the paper. Zhang gratefully acknowledges the kind support of Starting Research 
Fund of Nanjing Normal University and from the financial support of NSFC-11973029. This 
manuscript has made use of the data from the SDSS projects. The SDSS-III web site is 
http://www.sdss3.org/. SDSS-III is managed by the Astrophysical Research Consortium 
for the Participating Institutions of the SDSS-III Collaboration.

\section*{Data Availability}
The data underlying this article will be shared on reasonable request to the corresponding 
author (\href{mailto:xgzhang@njnu.edu.cn}{xgzhang@njnu.edu.cn}).

\label{lastpage}
\end{document}